\documentclass[referee]{aa} % for a referee version
\usepackage{psfig}
\usepackage{epsfig}
\usepackage{longtable}
\usepackage{lscape}

\def\kms{$\rm km\;s^{-1}$}

\def\FWHM{{\it FWHM}}

\begin{document}
\headnote{Research note}
\title{Spiral galaxies with a central plateau 
  in the gas velocity curve along the major axis\thanks{Based on observations 
  carried out at the European Southern Observatory (ESO 67.A-0286 and
  71.N-0229).}$^{\bf,}$\thanks{Table 4 is only available in
  electronic form at the CDS via anonymous ftp to cdsarc.u-strasbg.fr
  (130.79.128.5) or via http://cdsweb.u-strasbg.fr/cgi-bin/qcat?J/A+A/}} 

\author{L. Coccato \inst{1,2}       
	\and E. M. Corsini \inst{1}
	\and A. Pizzella \inst{1}
	\and F. Bertola\inst{1}}

\institute{
Dipartimento di Astronomia, Universit\`a di Padova, 
  vicolo dell'Osservatorio 2, 35122 Padova, Italy \and 
Kapteyn Astronomical Institute, Postbus 800,
  9700 AV Groningen, The Netherlands}

\offprints{L. Coccato, \\ e-mail: {\tt coccato@astro.rug.nl}}

%\date{\today}

\abstract{
We present the minor-axis kinematics of ionized gas and stars for a
sample of 5 spiral galaxies, which are characterized by either a zero
or a shallow gas velocity gradient along their major axis.
The asymmetric velocity profiles observed along the minor axis of NGC
4064 and NGC 4189 can be explained as due to the presence of a bar. 
This is also the case of NGC 4178 where the innermost portion of the
gaseous disk is nearly face on.
In NGC 4424 and NGC 4941, we measured non-zero gas velocities only in
the central regions along the minor axis and gas velocities drop to
zero at larger radii. This kinematic feature is suggestive of the
presence of an orthogonally-rotating gaseous component which is
confined in the innermost regions (i.e. an inner polar disk) to be
confirmed with integral-field spectroscopy.
\keywords{galaxies: kinematics and dynamics --- 
  galaxies: spirals --- galaxies: structure}
   }
\titlerunning{Spirals with a central plateau}
\authorrunning{L. Coccato et al.}
   \maketitle

\section{Introduction}
 
In a series of papers (Bertola et al. 1999; Sarzi et al. 2000; Corsini
et al. 2002, 2003; Coccato et al. 2004) we reported the case of spiral
galaxies whose major-axis velocity curves of the ionized-gas and/or
stellar component are characterized by a central plateau which is
extended a few arcseconds.
This uncommon feature has been also observed in
two-dimensional velocity fields (e.g., Sil'chenko \& Afanaziev 2004).
In some cases these galaxies exhibit a central velocity gradient along
the minor axis with non-zero velocities confined in the central
regions and dropping to zero at larger radii. In other cases the
minor-axis velocity gradient extends all over the observed range or is
missing (Coccato et al. 2004).

The presence of a major-axis velocity plateau together with a
minor-axis central velocity gradient is the most interesting case
since it constitutes the kinematic signature of an inner polar disk
(IPD hereafter). IPDs are small disks of gas and/or stars
($R\approx300$ pc), which are located in the center of lenticular and
spiral galaxies and are rotating in a plane perpendicular to that of
the main disk of their host. Most of these orthogonally-decoupled
structures have been discovered in last few years (see Corsini et
al. 2003 and references therein) and only a number of IPDs has been
studied in detail (Sil'chenko \& Afanaziev 2004).

Over the course of the last few years we have undertaken a program
aimed at detecting IPDs using long-slit spectroscopic observations
(Corsini et al. 2002, 2003; Coccato et al. 2004).
In the present paper, which is an extension of our previous studies on
IPDs, we analyzed a sample of 5 spiral galaxies, whose major-axis gas
rotation curve shows either a remarkable zero-velocity plateau (NGC
4424 as measured by Kenney et al. 1996 and Rubin et al. 1999, NGC 4941
as discussed in this paper) or a shallow velocity gradient (NGC 4064,
NGC 4178, NGC 4189 as found by Rubin et al. 1999).  NGC 4064, NGC
4178, NGC 4189, and NGC 4424 belong to the sample of spiral galaxies
measured by Rubin et al. (1999) to study the correlations between
kinematic disturbances, location in the cluster, and tidal encounters
of the galaxies in the Virgo cluster. NGC 4941 has been found by us in
a galaxy sample we studied not for this purpose. These galaxies are
promising candidates for our investigation beacuse of their major-axis
kinematic features. We therefore obtained spectra along their minor
axis in order to look for the possible presence of a velocity
gradient.
We present the ionized-gas and stellar kinematics of the sample
galaxies in Sect. \ref{sec:kinematics} and discuss our conclusions in
Sect. \ref{sec:conclusions}

\section{Ionized-gas and stellar kinematics}
\label{sec:kinematics}

The long-slit spectroscopic observations of the sample galaxies were
carried out with the New Technology Telescope (NTT) at the European
Southern Observatory (ESO) in La Silla (Chile) on May 16, 2001 (run
1), April 25-30, 2003 (run 2) and April 30, 2003 (run 3).  NTT mounted
the ESO Multi-Mode Instrument in red medium-dispersion spectroscopic
mode.
The grating No.~6 with 1200 grooves mm$^{-1}$ was mounted with the
No. 36 Tektronix TK2048 CCD in run 1 and with No.~62 MIT/LL 2048 CCD
in run 2. The grating No.~7 with 600 grooves mm$^{-1}$ was used in
combination with the No.~62 MIT/LL 2048 CCD in run 3. In run 1 each
$24\,\times\,24$ $\rm \mu m^2$ spectrum pixel corresponded to $\rm
0.32\AA\,\times\,0\farcs270$. In runs 2 and 3 a $2\times2$ binning was
applied and each $30\,\times\,30$ $\rm \mu m^2$ spectrum pixel
corresponded to $\rm 0.40\AA\,\times\,0\farcs332$ and $\rm
0.82\AA\,\times\,0\farcs332$, respectively. The slit was $1\farcs0$
wide and $5\farcm5$ long and gave an instrumental \FWHM\ of 1.1 \AA\
in run 1 (6400--7020 \AA) and 2 (4830--5470 \AA), and 2.2
\AA\ in run 3 (5850--7150 \AA). Typical seeing \FWHM\ was $1\farcs8$
in run 1 and $0\farcs9$ in runs 2 and 3 as measured by the ESO
Differential Image Motion Monitor.

For each sample galaxy, we took a spectrum with the slit along the
minor axis (NGC 4064 PA=60$^\circ$, NGC 4178 PA=120$^\circ$, NGC 4189
PA=175$^\circ$, NGC 4424 PA=5$^\circ$, and NGC 4941
PA=105$^\circ$). For NGC 4941 a major-axis spectrum (PA=15$^\circ$)
was obtained too. The position angles of the major and minor axis were
chosen according to RC3 and therefore relate to the orientation of the
outermost isophotes. Integration time of the galaxy spectra was
typically 5400 sec and split into different exposures to deal with
cosmic rays.  In run 2 we obtained spectra of some giant stars with
spectral type ranging from late-G to early-K to be used as templates
in measuring the minor-axis stellar kinematics of NGC 4064, NGC 4189,
and NGC 4941.  Arc lamp spectra were taken before and/or after every
object exposure to allow an accurate wavelength calibration.

\begin{figure*}[ht!]
\centering
\resizebox{\hsize}{!}{ 
\includegraphics[clip=true,angle=0]{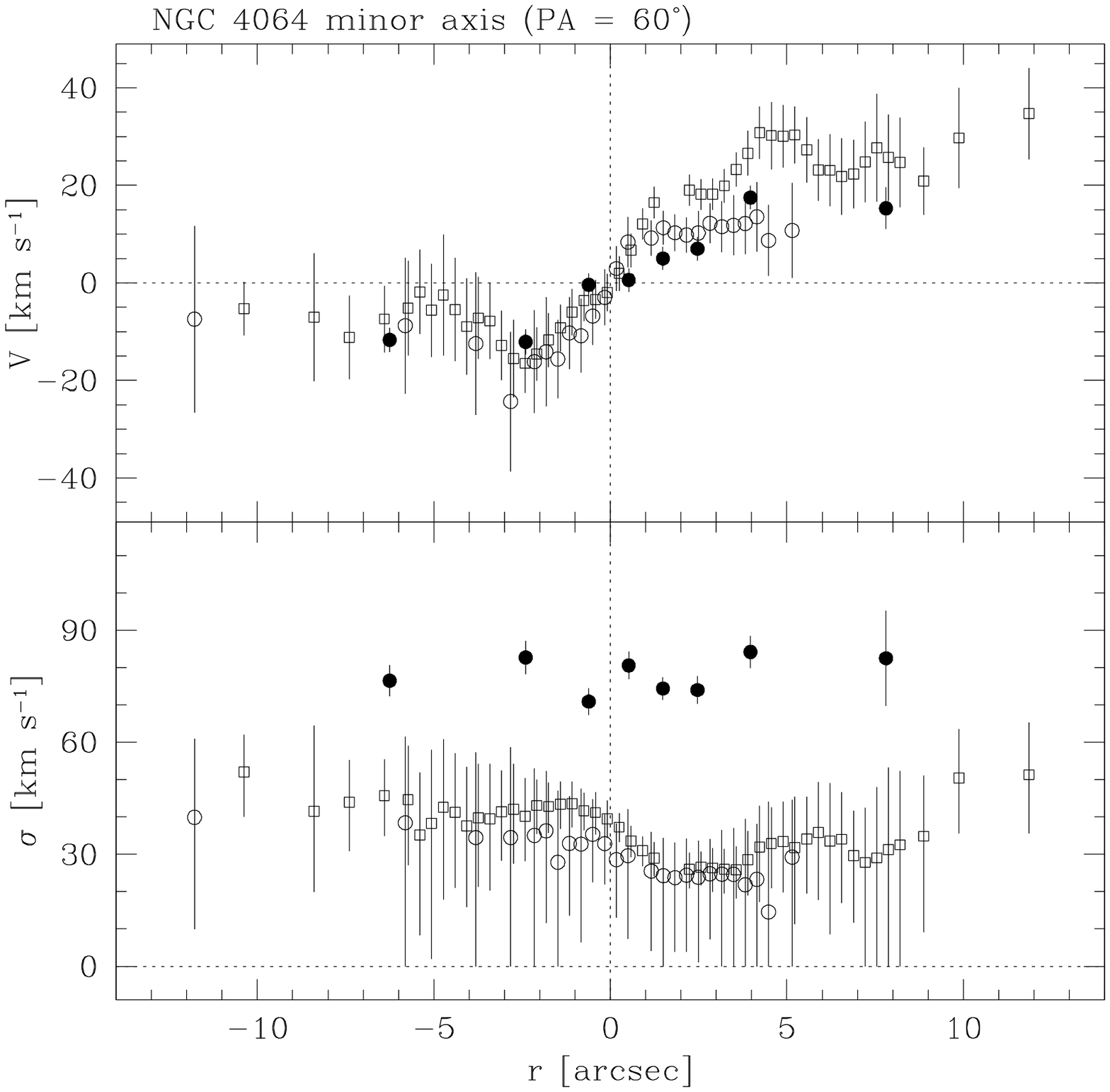} 
\includegraphics[clip=true,angle=0]{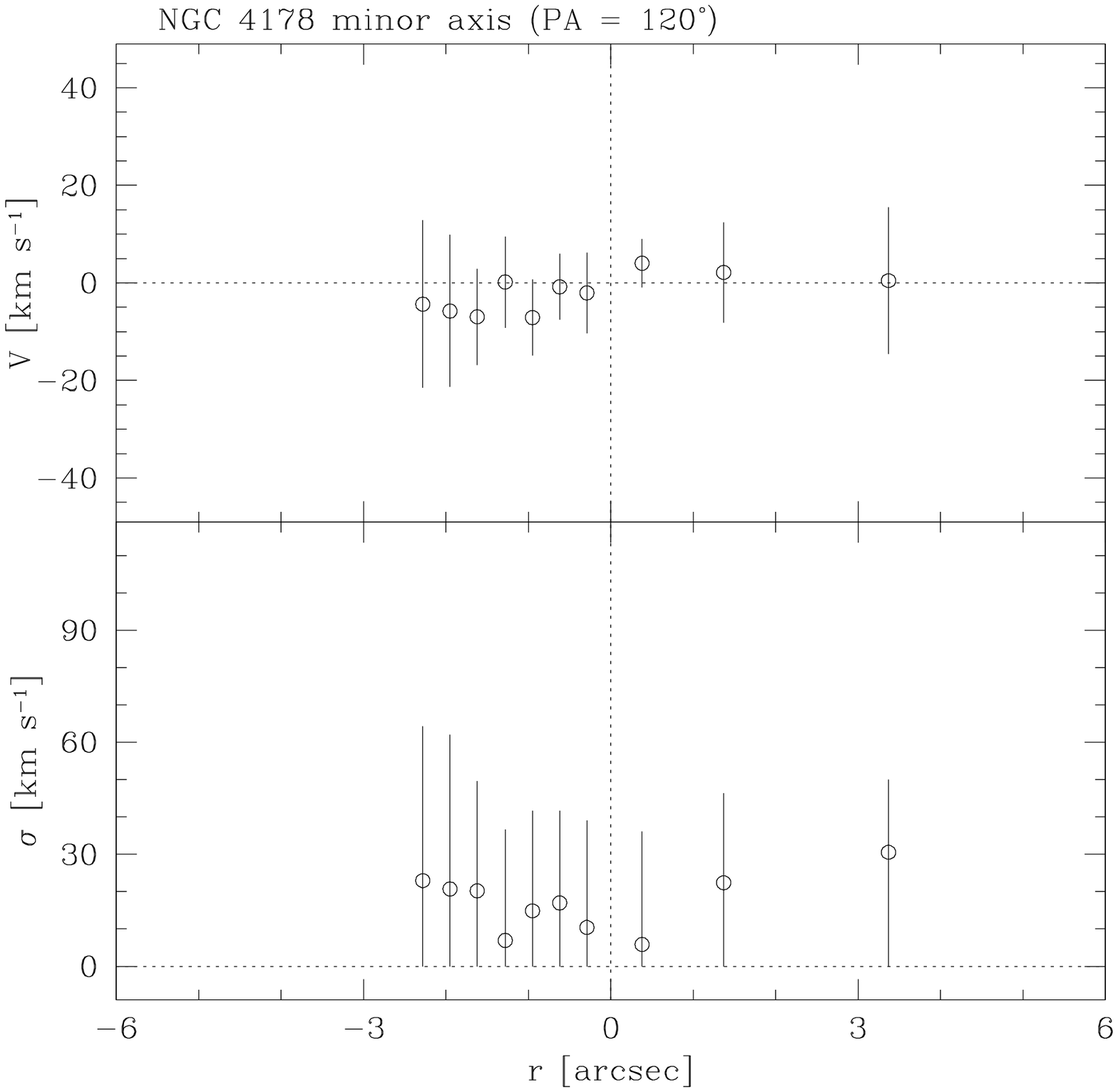}
\includegraphics[clip=true,angle=0]{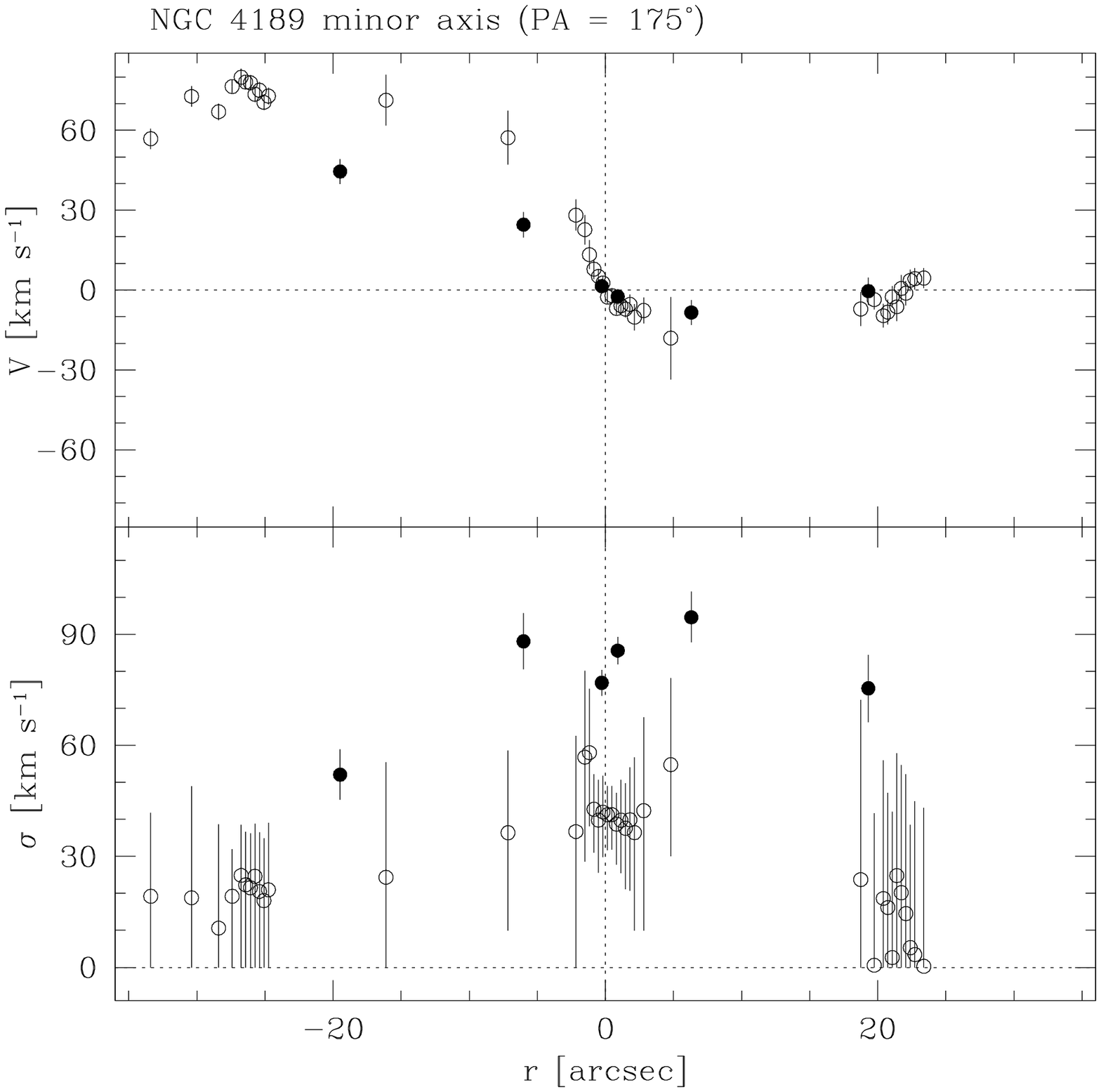} } 
\resizebox{\hsize}{!}{ 
\includegraphics[clip=true,angle=0]{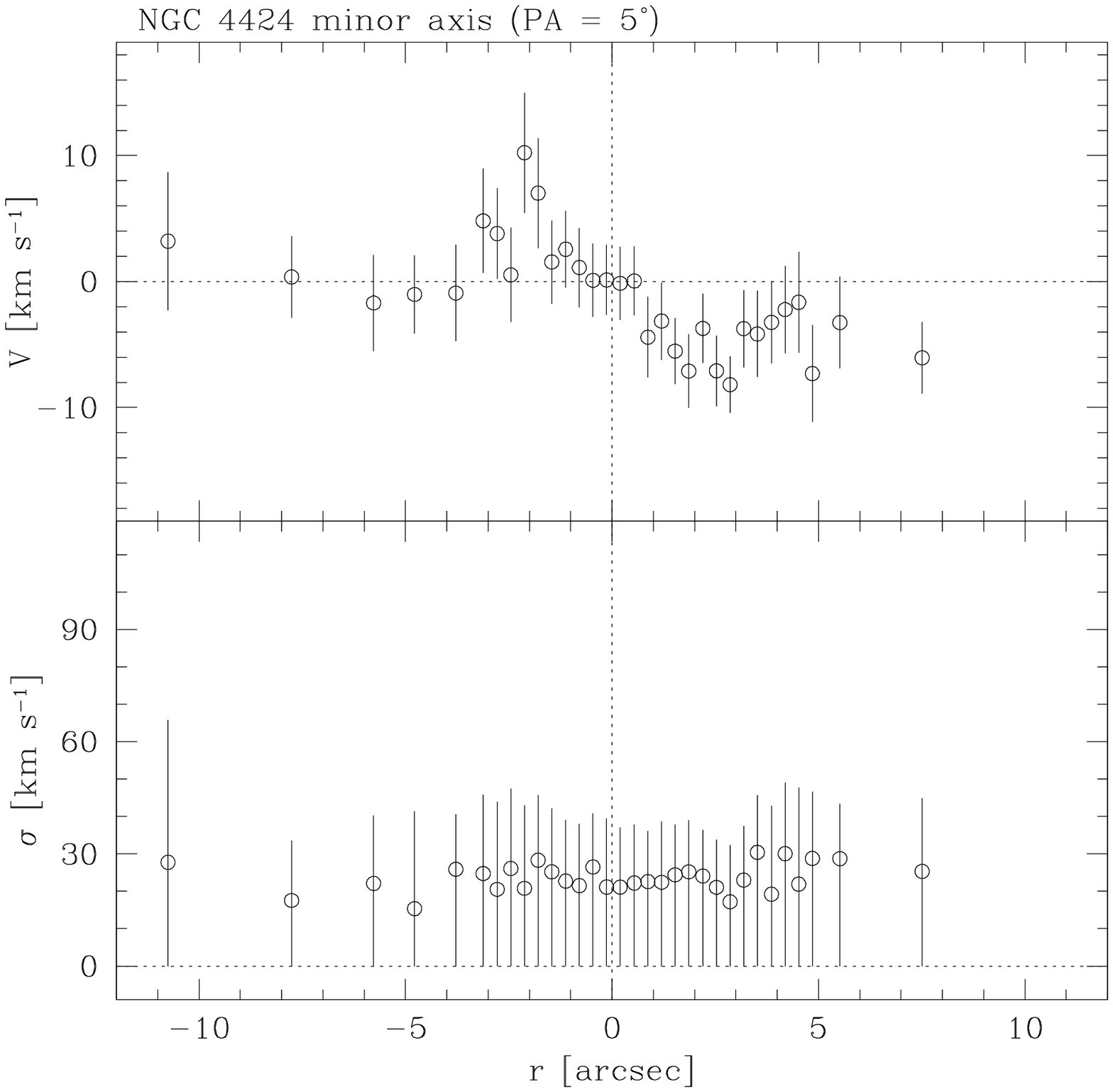} 
\includegraphics[clip=true,angle=0]{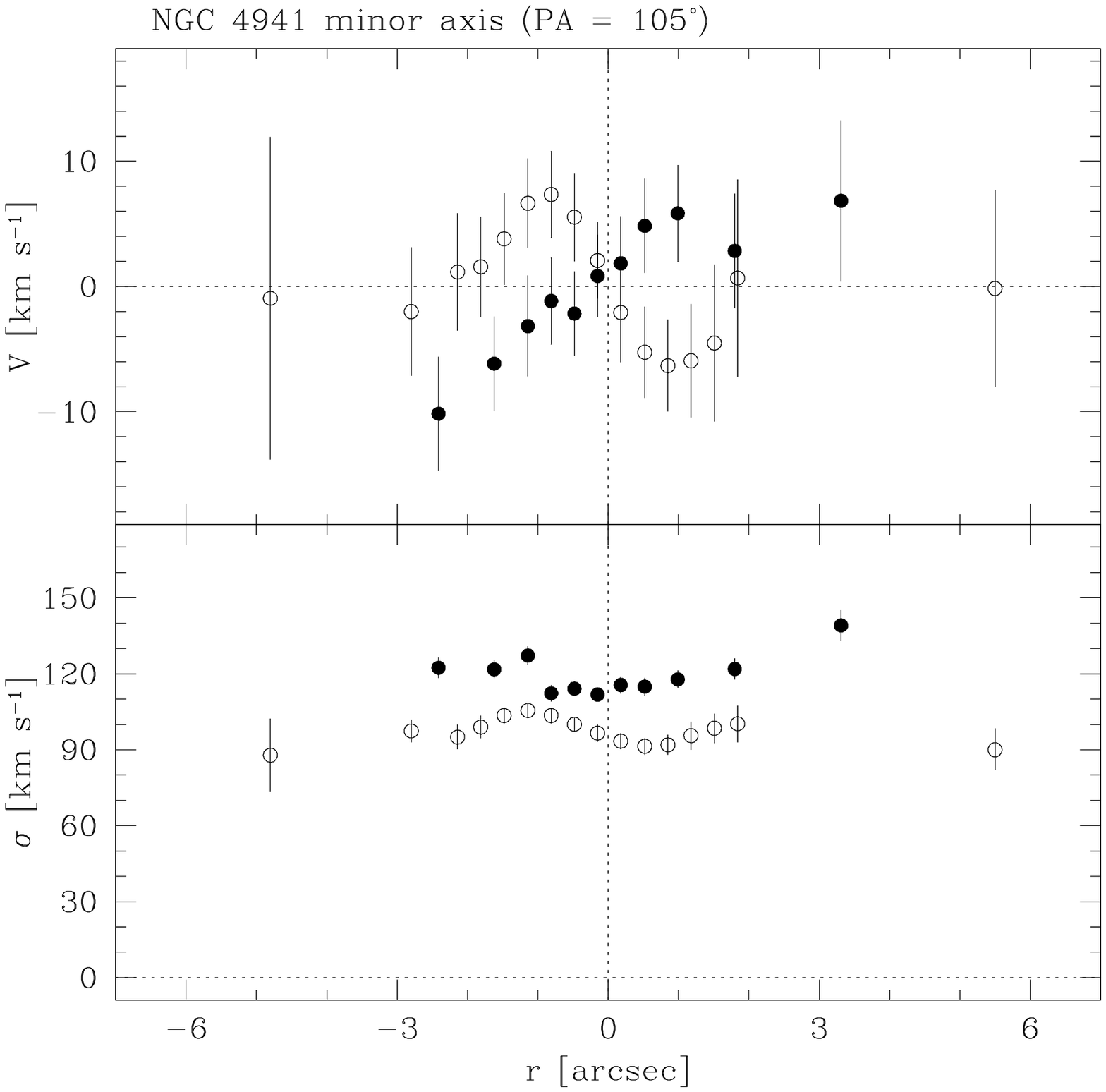}
\includegraphics[clip=true,angle=0]{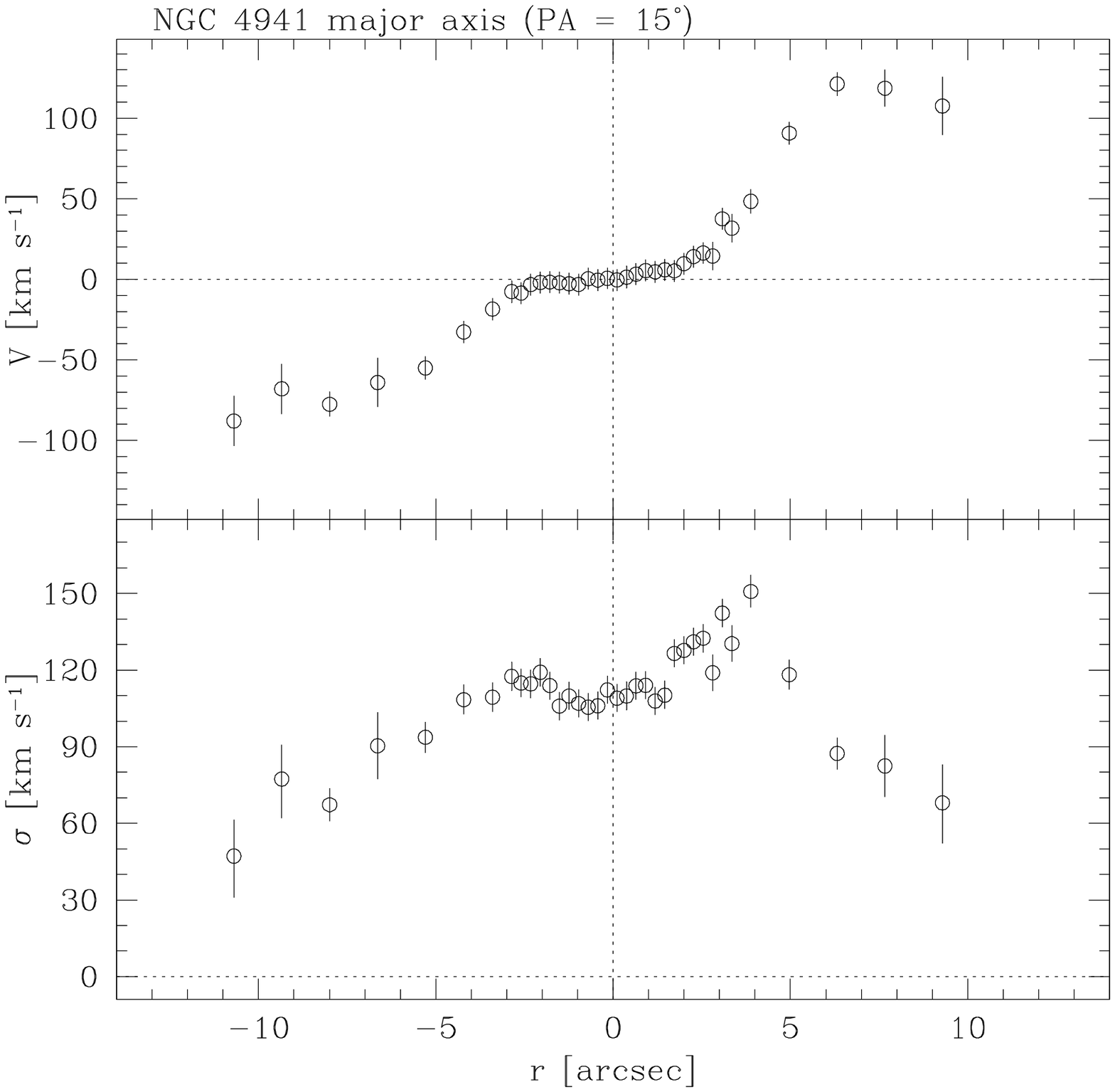}} 
\caption{Ionized-gas ({\it open symbols\/}) 
  and stellar ({\it filled circles\/}) kinematics measured along the
  optical minor axis of NGC 4064, NGC 4178, NGC 4189, NGC 4424, and
  NGC 4941, and the optical major axis of NGC 4941. For NGC 4064, the
  {\it open circles\/} and {\it open squares\/} refer to gas
  kinematics obtained in run 1 and 3, respectively.}
\label{fig:kinematics}
\end{figure*}

The basic data reduction and the measurement of the ionized-gas and
stellar kinematics was performed as in Corsini et al. (2003). 
The line-of-sight velocity and velocity dispersion profiles we
measured for the gaseous and stellar component of the sample galaxies
are presented in Fig. \ref{fig:kinematics} and values are reported in
Table 4. The line-of-sight velocities are the observed ones after
subtracting the systemic velocities derived as center of symmetry of
the gas velocity profile and without applying any correction for
galaxy inclination. We measured $cz=968\pm5$ \kms\ for NGC 4064,
$cz=390\pm5$ \kms\ for NGC 4178, $cz=2086\pm5$ \kms\ for NGC 4189,
$cz=454\pm5$ \kms\ for NGC 4424, $cz=1120\pm5$ \kms\ for NGC 4941.
The line-of-sight velocity dispersions are corrected for instrumental
velocity dispersion.

The following kinematic features are noteworthy in the galaxies of our
sample:

\medskip
\noindent
1. The major-axis velocity gradient of the ionized gas is zero in the
central regions of NGC 4189 (Rubin et al. 1999), NGC 4424 (Kenney
et al. 1996; Rubin et al. 1999), and NGC 4941
(Fig. \ref{fig:kinematics}). It is shallower than that measured at
larger radii in NGC 4064 and NGC 4178 (Rubin et al. 1999).
Further out the gas velocity remains low ($\la20$ \kms , NGC 4064) or
it increases to the last observed radius (NGC 4178, NGC 4189, NGC
4424, and NGC 4941).

\medskip
\noindent
2. Non-zero gas velocities are measured along the minor axis of all the
galaxies except for NGC 4178 (Fig. \ref{fig:kinematics}), in spite of
what it would be expected if the gas traced the circular velocity in
the disk plane.
The ionized-gas velocity profile measured along the optical minor axis
of NGC 4064 and NGC 4189 is strongly asymmetric. The gas velocity
rises to a maximum and it remains almost constant out to the last
observed point on the receding side. On the contrary, it drops to zero
at large radii on the approaching side.
The non-zero gas velocities are confined in the central regions along
the minor axis of NGC 4424 and NGC 4941, where the gas velocity curve
shows a steep gradient. The gas velocity increases to a maximum of
$\approx10$ \kms\ in the inner few arcsec, and further out it
decreases to zero. Our measurements along the minor axis of NGC 4424
are in agreement within errors with those by Kenney et al. (1996)
which are characterized by a lower spatial sampling.
A zero velocity gradient is observed for the ionized-gas component
along the minor axis of NGC 4178.
In all the galaxies the gas velocity dispersion remains low ($\la50$
\kms), except for NGC 4941 where it reaches $\approx100$
\kms\ in the inner $\sim5''$.

\medskip
\noindent
3. Non-zero velocities are measured along the minor axis of NGC 4064, NGC
4189, and NGC 4941 for the stellar component too. The stars rotate
slower than gas and have a larger velocity dispersion ($\approx80$
\kms ) in NGC 4064 and NGC 4189. Finally, in NGC 4941 stars are
counterrotating with respect to gas. They show a maximum observed
velocity of $\approx10$ \kms\ and a constant velocity dispersion of
$\approx 120$ \kms .

\section{Discussion and conclusions}
\label{sec:conclusions}

We have reported the ionized-gas and stellar kinematics along the disk
minor axis of 5 spiral galaxies, which are known to have either a zero
or a shallow gas velocity gradient along the disk major axis.

NGC 4064, NGC 4178, and NGC 4189 host a large-scale bar (RC3, but see
also CAG). Qualitatively, the minor-axis gas kinematics can be
explained as due to non-circular (e.g., Athanassoula 1992) or
off-plane (e.g., Friedli \& Benz 1993) motions induced on the gaseous
component by the tumbling triaxial potential of the bar. The innermost
portion of the gaseous disk of NGC 4178 is nearly face on since a zero
velocity gradient is observed along both major and minor axis. Bar can
account also for the stellar velocity gradient observed along the
minor axis of NGC 4064 and NGC 4189.

In spite of the morphological classification in RC3, NGC 4424 is an
unbarred galaxy. Kenney et al. (1996) first pointed out the presence
of strong non-circular gas motions and/or rotation in a plane
different from that of the outer stellar disk and attributed to a
recent merger. We suggest that the ionized-gas kinematics measured
for $|r|\la5''$ is consistent with the presence of an IPD.

No large-scale bar is observed in the early-type spiral NGC 4941
(Greusard et al. 2000), which is a Seyfert 2 galaxy (Stauffer 1982).
The peculiar gas kinematics can be associated to the extended
emission-line nuclear structure found by Martini et al. (2003), which
is roughly aligned with the galaxy minor axis and can be explained as
due to outflowing photoionized gas.
However, this does not preclude that such a gaseous component is an
IPD. It is worth noting that an IPD of gas and stars has been observed
in the Seyfert 2 galaxy NGC 4698 (Bertola et al. 1999; Bertola \&
Corsini 2000; Pizzella et al. 2002). Moreover NGC 4941 hosts a nuclear
bar, which is almost aligned with the galaxy major axis ($\rm \Delta
PA \approx 15^\circ$) and confined into the region where the kinematic
peculiarities are measured (Greusard et al. 2000).
Therefore the observed kinematics is consistent with the presence of
gas moving onto anomalous orbits in the tumbling potential of the
nuclear bar. This is also supported by the presence of gas in
retrograde motion relative to the stars along the disk minor axis
(i.e., in a direction close to the bar minor axis), where the
anomalous orbits are expected to be highly inclined with respect to
the rotation axis of the bar (see Friedli \& Benz 1993 for
details). In this scenario the gas is settled into a stable
configuration forming a strongly warped disk, whose innermost portion
corresponds to an IPD. This is the case of NGC 2217 (Bettoni et
al. 1990) which is characterized by the same gaseous and stellar
kinematics as NGC 4941.

We conclude that NGC 4424 and NGC 4941 are good candidates to be
followed up with integral-field spectroscopy in order to address the
presence of an IPD. These nuclear structures are remarkable enough to
require remarkable evidence and only the measurement of the
two-dimensional velocity field at high spatial resolution can fully
constrain the size and orientation of IPDs and will allow to get clues
about how they have been formed.

\end{document}